\documentclass{article}
\usepackage{amssymb}
\usepackage{amsfonts}

\input{tcilatex}

\begin{document}

\title{Quantum ground-state computation with static gates}
\author{Giuseppe Castagnoli\thanks{%
Information Technology Division \& Quantum Lab, Elsag spa, 16154 Genova,
Italy} \ \& David Ritz Finkelstein\thanks{%
School of Physics, Georgia Institute of Technology, Atlanta, GA 30332, USA}}
\maketitle

\begin{abstract}
We develop a computation model for solving Boolean networks that implements
wires through quantum ground-state computation and implements gates through
identities following from angular momentum algebra and statistics. The gates
are static in the sense that they contribute Hamiltonian 0 and hold as
constants of the motion; only the wires are dynamic. Just as a spin 1/2
makes an ideal 1-bit memory element, a spin 1 makes an ideal 3-bit gate.
Such gates cost no computation time: relaxing the wires alone solves the
network. We compare computation time with that of an easier Boolean network
where all the gate constraints are simply removed. This computation model is
robust with respect to decoherence and yields a generalized quantum speed-up
for all NP problems.
\end{abstract}

\section{ Introduction}

The prevailing approach to quantum computation evolved from classical
reversible algorithmic computation (Bennett 1979, Fredkin and Toffoli 1982),
where a stored program drives a sequence of elementary logically reversible
transformations. In reversible-algorithmic computation a time-varying
Hamiltonian drives a sequence of unitary transformations (Benioff 1982,
Feynman 1985). It was then found (first by Deutsch 1985) that entanglement,
interference and measurement yield in principle dramatic speed-ups over the
corresponding classical algorithms in solving some problems.

In spite of this important result, this form of computation faces two
possibly basic difficulties. Its speed-ups rely on quantum interference,
which requires computation reversibility. Decoherence may then limit
computation size below practical interest. Only two speed-ups of practical
interest have been found so far (factoring and database search), and none
since 1996.

Reversible-algorithmic computation is not the most general form of quantum
computation. Its limitations justify reconsidering quantum ground-state
computation (Castagnoli 1998, Farhi et al. 2001, Kadowaki 2002, among
others), a formerly neglected approach still believed to be mathematically
intractable. Quantum ground-state computation evolved from classical
ground-state computation (Kirkpatrick \& Selman 1994, among others), a
well-developed approach competitive with algorithmic computation for solving
Boolean networks. A Boolean network is a set of nodes (Boolean variables)
variously connected by gates and wires that impose relations on the
variables they connect (Fig. 1). A Boolean assignment satisfying all gates
and wires is a network solution. Roughly speaking, all NP problems can
readily be converted to the problem of solving a Boolean network.

In quantum ground-state computation, one sets up a quantum network whose
energy is minimum when all gates and wires are satisfied. In quantum
annealing, one form of ground-state computation, coupling the network with a
heat-bath of suitably decreasing temperature relaxes the network to its
ground state, a mixture of solutions (we assume with no significant
restriction that there is at least one). Measuring the node variables
(Hermitian operators with eigenvalues $0$ and $1$) yields a solution.

It is believed that quantum annealing yields a (still ill-defined) speed-up
over its classical counterpart. Quantum tunneling reduces the risk that the
network, in its way toward the absolute energy minimum, remains trapped in
local minima (e.g. Kadowaki 2002). However, long simulation times seriously
limit research on this approach.

Here we develop a hybrid mode of computation. We implement wires by
ground-state computation. We implement gates as algebraic identities
resulting from quantum symmetries and statistics.

We show that relaxation-computation time is comparable with that an easier
(more loosely constrained) logical network where all the gate constraints
implemented by quantum symmetries are removed. The comparison is based on a
special projection method. We show that the relaxation of the actual network
can be obtained as a special projection of the relaxation of the easier
comparison network. This projection method shortcuts mathematical complexity
and sheds light on the nature of this form of computation.

We conjecture that for this computation mode all hard-to-solve (NP) networks
become easy (P) and support this conjecture with plausible estimates.
Decoherence is not expected to be as serious a problem for this computation
model as for algorithmic computation since the network state is
intentionally a thermal mixture during most of the computation.

This discussion of quantum computation still belongs to the realm of
principles, like other literature on quantum ground-state computation, while
algorithmic-reversible computation is now almost a technology. Nevertheless
it is worth starting over with a new approach that might overcome
fundamental limitations of algorithmic computation.

\section{Computation model}

We use a network normal form composed just of wires and triodes (Fig.1).
Each triode $\tau $ --- properly a partial gate --- connects three nodes
labeled $\tau x$, $\tau y$, $\tau z$ (replaced by collective indices in
Fig.1) with the sum-2 relation 
\begin{equation}  \label{TRIODE}
{q}_{\tau x}+{q}_{\tau y}+{q}_{\tau z}=2,
\end{equation}
where ${q}$'s are Boolean variables and $+$ denotes arithmetical sum. The
three solutions are the rows of Table I.

Each wire $w(i,j)=w(j,i)$ is an equality relation ${q}_{i}={q}_{j}$ between
two nodes $i,j$ (Table II).

The example in Fig. 1, with $Q=6$ nodes, $W=4$ wires (lines), and $T=2$
triodes (dashed triangles), has just one solution: ${q}_{3}={q}_{5}=0$, ${q}%
_{1}={q}_{2}={q}_{4}={q}_{6}=1$.

\begin{center}
$%
\begin{array}[t]{ccc}
\begin{tabular}{|l|l|l|}
\hline
$q_{\tau{x}}$ & $q_{\tau{y}}$ & $q_{\tau{z}}$ \\ \hline
$0$ & $1$ & $1$ \\ \hline
$1$ & $0$ & $1$ \\ \hline
$1$ & $1$ & $0$ \\ \hline
\end{tabular}
& 
\begin{tabular}{|l|l|}
\hline
$q_{{i}}$ & $q_{{j}}$ \\ \hline
$0$ & $0$ \\ \hline
$1$ & $1$ \\ \hline
\end{tabular}
& 
\begin{tabular}{|l|l|l|c|}
\hline
$q_{\tau{x}}$ & $q_{\tau{y}}$ & $q_{\tau{z}}$ & {sy} \\ \hline
$0$ & $0$ & $0$ & $s$ \\ \hline
$0$ & $1$ & $1$ & $t$ \\ \hline
$1$ & $0$ & $1$ & $t$ \\ \hline
$1$ & $1$ & $0$ & $t$ \\ \hline
\end{tabular}
\\ 
\rule{0pt}{15pt} \mbox{Table I} & \mbox{Table II} & \mbox{Table III}%
\end{array}%
$

\[
\text{Fig. 1. A network}
\]
\end{center}

In the following we give idealized physical models of the various network
elements.

\subsection{Nodes and triodes}

Network nodes represent qubits, which here we consider in a most general way
as commuting Hermitian operators with eigenvalues $\ 0$ and $1$. In most
computation, the relations to be satisfied are achieved by a dynamical
development. In principle we may model any network relation by any valid
physical relation, however. Here we model the triode relation by a spin
identity. In the present computation model, each node belongs to a triode $%
\tau $, a spin 1 system, which might be two spins 1/2, \/ ${\frac{1}{2}}%
\sigma _{\tau 1},{\frac{1}{2}}\sigma _{\tau 2}$ \ in units $\hbar =1,$\ \ \
in a triplet state\/, with total spin vector $s_{\tau }={\frac{1}{2}}(\sigma
_{\tau 1}+\sigma _{\tau 2})$, $s_{\tau z}=\pm 1,0$. For each spin 1 we
define three qubits, 
\begin{equation}
q_{\tau x}=s_{\tau x}^{2},{\quad }q_{\tau y}=s_{\tau y}^{2},{\quad }q_{\tau
z}=s_{\tau z}^{2},
\end{equation}%
each representing a node of triode $\tau $.

By the composition of angular momentum, the three qubits of each proton pair
satisfy the XOR gate equation (Table III). The four rows of Table III
correspond to the singlet and the three triplet states of proton pair $\tau $%
, spanning the Hilbert space $\mathcal{H}_{\tau }^{(4)}$. We use \ $\mathcal{%
H}_{N}^{(4)}=\bigotimes_{\tau =1}^{T}\ \mathcal{H}_{\tau }^{(4)}$ as the
network space.

Note that network nodes are not divided into inputs and outputs of the
computation process as in algorithmic computation. They are all
simultaneously present in the network as commuting Hermitian operators,
related by time-independent identities that we use as gates.

To simplify the physical model, we assume that the spatial wave function of
each proton pair $\tau $ is frozen throughout the computation in a stable
antisymmetric ground state, like that of the protons of an orthohydrogen
molecule. In the triplet state the three qubits of any triode $\tau$ obey
the triode sum-2 relation of (\ref{TRIODE}) and Table I. This follows from
angular momentum composition and triplet symmetry, both extradynamical
relations; the triode Hamiltonians are zero in $\mathcal{H}_{N}^{(4)}$.

Let $\mathcal{H}_{\tau }^{(3)}$ be the space spanned by the three triplet
states of triode $\tau $. \ \ $\mathcal{H}_{N}^{(3)}=\bigotimes_{\tau
=1}^{T}\ \mathcal{H}_{\tau }^{(3)}\subset \mathcal{H}_{N}^{(4)}$ is the
network subspace with all the\ triodes satisfied.

\subsection{Wires}

We define the frustration Hamiltonian of wire $w(i,j)$ in $\mathcal{H}%
_{N}^{(4)}$ by $H_{i,j}^{(4)}=g\left( q_{i}-q_{j}\right) ^{2},$ where $g$ \
is a coefficient to provide the dimension of energy. The eigenvalues of $%
H_{i,j}^{(4)}$ are zero when the wire is satisfied, $g$ when it is not (the
wire is then \textquotedblleft frustrated\textquotedblright ). All $%
H_{i,j}^{(4)}$ commute. Therefore the network frustration Hamiltonian is 
\begin{equation}
H_{N}^{(4)}=g\sum_{\{w\}}\left( q_{i}-q_{j}\right) ^{2},
\end{equation}%
where $\{w\}$ is the set of all wires in the network.

Let $X_{\tau }$ be the exchange operator for the two protons of triode $\tau 
$. $H_{N}^{(4) }$ is symmetric under all the $X_{\tau }$ ($X_{\tau }$ $%
H_{N}^{(4) }=H_{N}^{(4) }X_{\tau }$), since the $q$'s are. Therefore the
triplet symmetry projection operator $T_{\tau}$ of the triode $\tau$ is a
constant of motion of $H_{N}^{(4) }$. If the initial network state is in $%
\mathcal{H}_{N}^{(3)}$, under $H_{N}^{(4) }$ it remains in it. The ground
state of $H_{N}^{(4) }$ (in $\mathcal{H}_{N}^{(3)}$) hosts a mixture of
network solutions, since all wires and triodes are satisfied.

\subsection{Ising model}

$H_{N}^{(4)}$ is quadrilinear in the spin components $s_{\tau x,y,z}$\/. As
a step toward implementation, we show that $H_{N}^{(4)}$ can be represented
by pairwise spin-spin interactions by adjoining two \textquotedblleft
idlers,\textquotedblright\ auxiliary spin-$1/2$ variables $\sigma
_{i},\sigma _{j}$, to the spin-1 variables $s_{i},s_{j}$ already defined for
each wire $w(i,j)$. For convenience we normalize the idler variables to
eigenvalues $\sigma _{i}=0,1$. One of many suitable frustration Hamiltonians
for wire $w(1,2)$, connecting the nodes $q_{1}=s_{1}^{2}$ and $%
q_{2}=s_{2}^{2}$ (using collective indices 1, 2) is the bilinear form: 
\begin{equation}
H_{w(1,2)}=g\left[ \left( s_{1}+s_{2}\right) ^{2}+5\left( s_{1}+s_{2}\right)
\left( -\sigma _{1}+\sigma _{2}\right) +\sigma _{1}\sigma _{2}+6(\sigma
_{1}+\sigma _{2})\right] .
\end{equation}%
This is chosen so that (as is readily checked) the ground-state projection
operator of the wire (= nodes + idlers) $\rho _{0}(s_{1},s_{2},\sigma
_{1},\sigma _{2})$ has energy eigenvalue 0 and includes eigenvectors $%
\left\vert s_{1},s_{2},\sigma _{1},\sigma _{2}\right\rangle $ with all five
combinations of spin-1 eigenvalues that satisfy the wire, namely $(s_{1}=\pm
1,s_{2}=\pm 1)$ and $(s_{1}=0,s_{2}=0)$, with correlated idler eigenvalues $%
s_{1},s_{2}$. All the states where the wire is frustrated ($s_{1}=\pm 1$, $%
s_{2}=0$; $s_{1}=0$, $s_{2}=\pm 1$) have energy $\geqslant g$ for any values
of $\sigma _{1},\sigma _{2}$\/.

The reduced state of the nodes in the ground state is the trace over the
idlers, 
\begin{equation}
\rho_0(s_1, s_2)= \mbox{tr}_{\sigma_1, \sigma_2} \rho_0(s_1, s_2, \sigma_1,
\sigma_2)
\end{equation}
Since all the wires are satisfied in the network state $\rho_0(s_1, s_2,
\sigma_1, \sigma_2)$, they are satisfied in the reduced node state $%
\rho_0(s_1, s_2)$. After the network relaxes to the ground state, we can
find a solution to the network problem by simultaneously measuring both node
bits $q_1, q_2$, ignoring the idlers.

It is then straightforward to construct the network purely out of spins 1/2
with pairwise coupling, as in the Ising model. We leave the idlers alone but
replace each spin 1 by the sum of two spins 1/2 with a coupling that favors
the triplet (parallel) state over the singlet (antiparallel) overwhelmingly.

\subsection{Heat-bath and coupling}

It is convenient to use heat-bath quanta that are distinguishable from the
network quanta. We use a photon-filled cavity with Hilbert space $\mathcal{H}%
_{B}$. $\mathcal{H}_{B}^{(4) }:=\mathcal{H}_{N}^{(4)}\otimes \mathcal{H}_{B}$
is the \textquotedblleft system\textquotedblright\ ($=$network$+$bath) {%
space,\ }$\mathcal{H}^{\left( 3\right) }:=\mathcal{H}_{N}^{(3)}\otimes 
\mathcal{H}_{B}$ is the subspace with triplet symmetry, all triodes
satisfied.

We denote by $H_{B}^{(4)}(t)$ the heat-bath Hamiltonian and define the
network-bath coupling in $\mathcal{H}^{(4)}$ by 
\begin{equation}
{H}_{I}^{(4)}(t)=g\sum_{\tau }\left[ \vec{B}_{\tau }(t)\cdot \vec{\sigma}%
_{\tau 1}+\vec{B}_{\tau }(t)\cdot \vec{\sigma}_{\tau 2}\right] .
\end{equation}%
This couples each proton spin to the small random Gaussian time-varying
magnetic field $\vec{B}_{\tau }(t)$ of the photon field at the site of the
spin. To maintain the symmetry that we have assumed, we assume that the two
protons of the same triode $\tau $ experience the same magnetic field and
that the spatial wave functions of different proton pairs do not overlap.

Therefore triplet symmetry (satisfaction of all triodes) is a constant of
motion of ${H}_{I}^{(4) }(t)$\/, thus also of the system Hamiltonian ${H}%
^{(4) }(t) =H_{N}^{\left( 4\right) }+H_{B}^{(4) }(t) +{H}_{I}^{(4) }(t) $;
in fact $H_{N}^{(4) }$ is already symmetric.

\subsection{Network relaxation process}

With a suitable time-variation of $\vec{B}_{\tau }(t)$, ${H}^{(t) } $
relaxes the network to its zero point.

Let $\left\vert \psi ,t\right\rangle $ be the state of the system at time $t$%
. The development of $\left\vert \psi ,t\right\rangle $ is generated by ${H}%
^{(4) }(t) $ according to the Schr\"{o}dinger equation. The relaxation of
the network state is described by the statistical operator $\rho _{N}(t):=%
\mbox{tr}_{B}\left( \left\vert \psi ,t\right\rangle \left\langle \psi
,t\right\vert \right) $, where tr$_{B}$ means trace over the heat-bath
degrees of freedom. If $\rho _{N}(t)$ starts in $\mathcal{H}_{N}^{(3)}$ it
remains in it, since both the wire frustration Hamiltonian and the
heat-bath\ coupling are symmetric.

A direct estimate of relaxation time is likely mathematically intractable,
and a simulation is very long. We take a shortcut that also sheds light on
the nature of this hybrid computation.

We compare the network relaxation time with that of an easier network
obtained by replacing all triodes (Table I) by XOR gates (Table III): as if
proton indistinguishability were suspended -- each proton pair were replaced
by a deuteron.

The restriction to $\mathcal{H}^{(3)}$ vanishes: a network of XOR gates and
wires is loosely constrained and easy to solve. In particular ${q}_{i}= 0$
for all $i$ is always a solution. A XOR network is solvable in poly($Q$)
time in classical computation and, one reasonably supposes, in the present
hybrid computation also. Note that the XOR gates of the comparison network
are also extra-dynamical; they represent a physical law, namely the
composition of angular momentum.

That the relaxation of the comparison network is quick can be plausibly seen
as follows.

We first replace $H_{N}^{(4) }$ by the new Hamiltonian 
\begin{equation}
H_{N}^{\prime (4) }=H_{N}^{(4) }\left[ 1+\frac{g^{\prime }}{g}\sum_{\tau
}\left( q_{\tau x}^{2}+q_{\tau y}^{2}+q_{\tau z}^{2}\right) \right] .
\end{equation}
Since each triode has exactly two nodes equal to $1$ (Table I), we have $%
H_{N}^{\prime (4) }=H_{N}^{(4) }\left(1+2T\frac{g^{\prime }}{g}\right) $. In
the case of the actual network we have merely multiplied $H_{N}^{(4) }$ by a
constant factor and the replacement is inessential. The ground states of $%
H_{N}^{(4) }$ and $H_{N}^{\prime (4) }$\ are the same, and the respective
energy landscapes are proportional.

Not so for the comparison network, no longer restricted to $\mathcal{H}%
_{N}^{(3)}$. If $g^{\prime }\gg g$, the energy \ landscape of $H_{N}^{\prime
(4) }$ has a gradient everywhere toward the solution $q_{i}=0$ allowed by
Table III. There are no local minima that can trap the comparison network on
its way toward the absolute minimum; thus the relaxation time of the
comparison network is reasonably poly($Q$).

We conjecture that introducing $H_{N}^{\prime (4) }$ is unnecessary. The XOR
gates, being extra-dynamical, would not affect relaxation time; they could
be removed, which would leave us with a set of independently relaxing wires,
and no local minima.

\subsection{Comparison system}

The asymmetric Hamiltonian of the comparison system in $\mathcal{H}^{\left(
4\right) }$ is $H^{\mathrm{A}}(t)=H_{N}^{(4)}+H_{B}^{(4)}(t)+H_{I}^{\mathrm{A%
}}\left( t\right) $, where 
\begin{equation}
H_{I}^{\mathrm{A}}(t)=g\sum_{\tau }\left[ \vec{B}_{\tau 1}(t)\cdot \vec{%
\sigma}_{\tau 1}+\vec{B}_{\tau 2}(t)\cdot \vec{\sigma}_{\tau 2}\right]
\end{equation}%
is the asymmetric coupling. Now we have two independent random Gaussian
time-varying magnetic fields at each proton site, such that%
\begin{equation}
\vec{B}_{\tau }(t)=\left[ \vec{B}_{\tau 1}(t)+\vec{B}_{\tau 2}(t)\right] /2
\end{equation}%
is the actual heat-bath. This is always possible since the sum of two
Gaussian distributions is also Gaussian.

Let $\left\vert \varphi ,t\right\rangle $ be the state of the comparison
system, whose\ development is generated by ${H}^{\mathrm{A} }\left( t\right) 
$. The comparison network relaxation is described by the statistical operator%
$\ \rho _{N}^{\mathrm{A} }(t):=\mbox{tr}_{B}\left( \left\vert \varphi
,t\right\rangle \left\langle \varphi ,t\right\vert \right) $. \ 

\subsection{Continuous projection method}

The symmetrization operator for all the proton pairs is 
\begin{equation}
P:=\prod_{\tau =1}^{T}{\frac{1+X_{\tau }}{2}}.
\end{equation}%
It projects $\mathcal{H}^{(4)}$ on $\mathcal{H}^{(3)}.$ Clearly $PH_{I}^{%
\mathrm{A} }(t) P=H_{I}^{(4)}(t)$\/, and so 
\begin{equation}
PH^{\mathrm{A} }(t) P=H^{(4)}(t),
\end{equation}
given that $H_{N}^{(4)}$is symmetric and $P$ is the identity in $\mathcal{H}%
_{B}$.

The development of the actual system (hard triode network and bath) in $%
\mathcal{H}^{(3)}$ is driven by $H^{(4)}(t)$; that of the comparison system
(easy XOR network and bath) in $\mathcal{H}^{(4)}$ is driven by $H^{\mathrm{A%
}}(t)$\/. We show that, given (11), the continuous projection on $\mathcal{H}%
^{(3)}$ of the development of the comparison system yields the development
of the actual system.

Let $\left\vert \psi ,t\right\rangle $ $\ $be an initial state of the actual
system in $\mathcal{H}^{(3)}$, therefore $P\left\vert \psi ,t\right\rangle
=\left\vert \psi ,t\right\rangle .$ Under $H^{(4)}(t) ,$ it develops into%
\begin{equation}
\left\vert \psi ,t+dt\right\rangle =\left( 1-iH^{(4)}(t) dt\right)
\left\vert \psi ,t\right\rangle .
\end{equation}%
Under $H^{\mathrm{A} }(t) $, it develops into%
\begin{equation}
\left\vert \varphi ,t+dt\right\rangle :=\left( 1-iH{^{\mathrm{A} }}\left(
t\right) dt\right) \left\vert \psi ,t\right\rangle ,
\end{equation}%
in general non-symmetric. We restore particle indistinguishability by
projecting $\left\vert \varphi ,t+dt\right\rangle $ on $\mathcal{H}^{(3)}$,
symmetrizing it: 
\begin{equation}
P\left\vert \varphi ,t+dt\right\rangle =\left( P^{2}-iPH{^{\mathrm{A} }}(t)
P\right) \left\vert \psi ,t\right\rangle =\left( 1-iH^{(4)}(t) dt\right)
\left\vert \psi ,t\right\rangle ={}\left\vert \psi ,t+dt\right\rangle .
\end{equation}%
We can see that the continuous projection of the comparison development
yields the actual development.

\subsection{Comparing computation times}

Computation time is by assumption poly($Q$) for the comparison easy XOR
network. To estimate that of the actual hard triode network, we decompose a $%
\Delta T$ into $N=\Delta T/\Delta t$ \ consecutive\ time slices $\Delta
t_{i} $ $\equiv \left[ t_{i},t_{i+1}\right] $ of equal length $\Delta t.$
Within each $\Delta t_{i},$ we consider the\ relaxation of the comparison
XOR network in $\mathcal{H}^{(4)}$, described by $\rho _{N}^{\mathrm{A} }(t) 
$. At the end of each $\Delta t_{i},$\ we project\textbf{\ }$\rho _{N}^{%
\mathrm{A} }(t) $\textbf{\ } on $\mathcal{H}^{(3)} $, then take the limit $%
\Delta t\rightarrow 0$. This yields the actual network relaxation $\rho
_{N}(t) $.

Within each $\Delta t_{i}$ we consider the decomposition%
\begin{equation}
\rho _{N}^{\mathrm{A} }(t) :=\rho _{0}(t)+\rho _{F}(t)+\rho _{V}(t)\/.
\end{equation}

\begin{itemize}
\item $\rho _{0}(t)$ describes networks with satisfied triodes and wires,
namely solutions of the actual network; its probability is $p_{0}(t):=%
\mbox{tr}\rho _{0}(t) .$

\item $\rho _{F}(t)$ describes networks with satisfied triodes and at least
one frustrated wire; $p_{F}(t):=\mbox{tr}\rho _{F}(t) .$

\item $\rho _{V}(t)$ describes networks with at least one violated triode,
wires are either satisfied or frustrated; $p_{V}(t):=\mbox{tr}\rho
_{V}\left( t\right) .$
\end{itemize}

We have considered all the possible states of the comparison network.
Therefore $p_{0}(t)+p_{F}(t)+$ $p_{V}(t)=1$. $p_{V}(t)$ goes to zero with $%
\Delta t$ and is annihilated by each projection.

The actual network-bath interaction soon randomly generates a $\rho
_{0}(t_{h})$, a mixture of solutions of the actual network, with an
extremely small probability $p_{0}(t_{h})$ $=O\left( 1/2^{Q}\right) $. For a
given confidence level, $t_{h}$ does not depend on $Q$.

For $t>t_{h}$ we apply the projection method. $p_{0}(t_{h})=O\left( 1/2^{Q{\ 
}}\right) $ becomes the nucleus of condensation of the network solutions.

Within each and every $\Delta t_{i}$, we take a constant-average logarithmic
rate of decrease $k$ of the frustration energy of the comparison network: 
\begin{equation}
E_{N}(t_{i+1})=\left( 1-k\Delta t\right) E_{N}(t_{i}).
\end{equation}%
We will show later that there is no error in taking a constant-average rate.
The relaxation time constant $1/k$ is by assumption poly($Q$).

We have $E_{N}(t):=$\mbox{tr}$\rho _{N}^{\mathrm{A}}(t)H_{N}^{(4)}=\mbox{tr}%
\rho _{F}(t)H_{N}^{(4)}$. In fact there is no contribution from $\rho _{0{\ }%
}(t)$, which is the ground state of $H_{N}^{(4)},$ and a possible
contribution from$\ \rho _{V}(t)$ would anyhow be second order
infinitesimal. Let $p_{F}^{\left( j\right) }(t)$\ be the $j$-th (population)
element of the diagonal of $\rho _{F}(t)$. Of course \ 
$\sum\limits_{j}p_{F}^{\left( j\right) }(t)=\mbox{tr}\rho _{F}(t)=$ $%
p_{F}(t) $. $H_{N}^{\left( 4\right) }$ is diagonal, thus we have $E_{N}(t)%
=\sum\limits_{j}$\ $p_{F}^{\left( j\right) }(t)E^{\left( j\right) }$, where $%
E^{\left( j\right) }$\ is the $j$-th diagonal element of $H_{N}^{(4)}$.\
Therefore $E_{N}(t)$ and $p_{F}(t)$ go to zero together. Thus on average: 
\begin{equation}
p_{F}(t_{i+1})=\left( 1-k\Delta t\right) p_{F}(t_{i}).
\end{equation}%
The decrease of $p_{F}(t)$ implies an equal increase of $p_{0}(t)$ $+$ $%
p_{V}(t).$ It is reasonable and conservative to consider the increase of $%
p_{V}(t)$ dominant. In fact the relaxation of the comparison network is
quicker because triodes can be violated.

Note that we compare relaxation rates, not directions: the comparison
network can head toward $\mathcal{H}_{N}^{(4)}\sim \mathcal{H}_{N}^{(3)}$,
the actual network remains in $\mathcal{H}_{N}^{(3)}$. It is like comparing
the speed of the keel and the wind on a broad reach. Speeds are
proportional, while keel and wind go to different places.

Furthermore, $H_{N}^{(4) }$ does not couple $\rho _{0}(t)$ with $\rho
_{F}(t) $ or $\rho _{V}(t)$. In fact $H_{N}^{(4) }\rho _{0}(t)$ = $\rho
_{0}(t)H_{N}^{(4) }=0$. Therefore $p_{0}(t)$ neither decreases nor increases
on average. Since $p_{F}(t)$ decreases and $p_{0}(t)$ does not, the ratio $%
p_{F}(t)/p_{0}(t)$ decreases. When we project on $\mathcal{H}_{N}^{(3)}$ at
the end of $\Delta t_{i}$, we remain with a smaller $p_{F}(t)$ and a\ larger 
$p_{0}(t)$ (probability of solutions of the actual network). \ 

We can focus on the \textquotedblleft take-off\textquotedblright\ of the
probability of solution from the extremely small value $p_{0}(t_{h})=O\left(
1/2^{N}\right) $ to $p_{0}(t)$ close to $1$, say $p_{0}(t)=1/10$.

During take-off and within each $\Delta t_{i}$ we have $p_{F}(t)\approx 1$; $%
\ p_{V}(t)$ grows from $p_{V}(t_{i})=0$ to $p_{V}(t_{i}+\Delta t)=k\Delta
tp_{F}(t_{i})\approx k\Delta t$, because of (17) and the assumption that $%
p_{0}(t)$ remains unaltered. The projection at the end of $\Delta t_{i}$
annihilates $p_{V}(t_{i}+\Delta t)$ reducing $\rho _{N}^{\mathrm{A}}(t)$ by
about $k\Delta t.$ Renormalizing $\rho _{N}^{\mathrm{A}}(t)$\ then
multiplies $p_{0}(t)$ by about $\left( 1-k\Delta t\right) ^{-1}\approx
1+k\Delta t$ at each $\Delta t_{i}$. After a time $\Delta T=N\Delta t$ and
in the limit $\Delta t\rightarrow 0$, we obtain for the actual network: 
\begin{equation}
p_{0}(t_{h}+\Delta T)\approx p_{0}(t_{h})\lim_{\Delta t\rightarrow 0}\left(
1+k\Delta t\right) ^{\frac{\Delta T}{\Delta t}}=p_{0}(t_{h})e^{k\Delta
T}\approx \frac{1}{2^{Q}}e^{k\Delta T}.
\end{equation}%
The probability of having solutions of the actual network becomes $O\left(
1\right) $ in a time $\Delta T\approx Q/k=Q$poly($Q$)$=$poly($Q$).

Using a different $k_{i}$ for each $\Delta t_{i}$, with average value $k={%
\sum_{i}k_{i}\Delta t_{i}/\Delta T}$, yields the same result: $e^{k\Delta T}$
in $\left( 18\right) $\ should be\ replaced by $\prod_{i}e^{k_{i}\Delta
t_{i}}=e^{k\Delta T}$.

\section{Conclusions}

The extradynamical algebraic relations expressing particle statistics and
angular momentum composition can replace the dynamical algebraic relations
following from equations of motion as computational gates. In this new form
of quantum computation, the gates of a Boolean network are always satisfied
as constants of the motion, leaving only equality relations (wires) to be
implemented dynamically. This form of quantum computation is expected to be
robust, since it relies on thermal mixtures, not pure states, and is
plausibly conjectured to be fast, turning all NP problems in principle into
P. As in quantum algorithmic computation, the speed-up is due to the
extradynamical character of the computation (Castagnoli \& Finkelstein 2001,
2002).

This model of computation highlights the conceptual difference between how
structures can be assembled in the classical and quantum domain. Quantally
it is as though one could assemble a jigsaw puzzle simply by piling the
pieces up and letting gravity lower them into mutual positions that solve
the puzzle, analogously to quantum wire relaxation. Classically this way of
assembling the pieces would be plagued by local energy\ minima. One may
wonder whether the assembly of biological molecules under hydrophobic
pressure draws on similar quantum effects.

Many of the ideas propounded in this work were developed through discussions
with Artur Ekert.

\bigskip

\noindent{\Large \textbf{References}}

\bigskip

\noindent Benioff, P. 1982 \textit{Phys. Rev. Lett.} \textbf{48},1581.

\noindent Bennett, C. 1979, Logical Reversibility of Computation, \textit{%
IBM J. Res. Dev.} \textbf{6}, 525.

\noindent Castagnoli, G. 1998 \textit{Physica D }\textbf{120,} 48.

\noindent Castagnoli, G. \& Finkelstein, D. 2001 \textit{Proc. R. Soc. Lond. 
}A \textbf{457}, 1799.

\noindent Castagnoli, G. \& Finkelstein, D., Quantum-Statistical
Computation, arXiv:quant-ph/0111120 v4 30 Jan 2002.

\noindent \noindent Deutsch, D. 1985 \textit{Proc. R. Soc. Lond. }A \textbf{%
400,} 97.

\noindent Farhi, E. Goldstone, J., Gutmann, S., Lapan, J., Lundgren, A. \&
Preda, D. 2001 \textit{Science} \textbf{292}, 472\ 

\noindent Feynman, R. 1985 \ \textit{Opt. News} \textbf{11}, 11.

\noindent Fredkin, E. \&\ Toffoli, T. 1982 \textit{Int. J. Theor. Phys.} 
\textbf{21}, 219.

\noindent Kadowaki, T., Study of Optimization Problems by Quantum Annealing,
arXiv:quant-ph/0205020 v1 5 May 2002.

\noindent Kirkpatrick, S. \& Selman, D. 1994 \textit{Science} \textbf{264},
1297.

\noindent

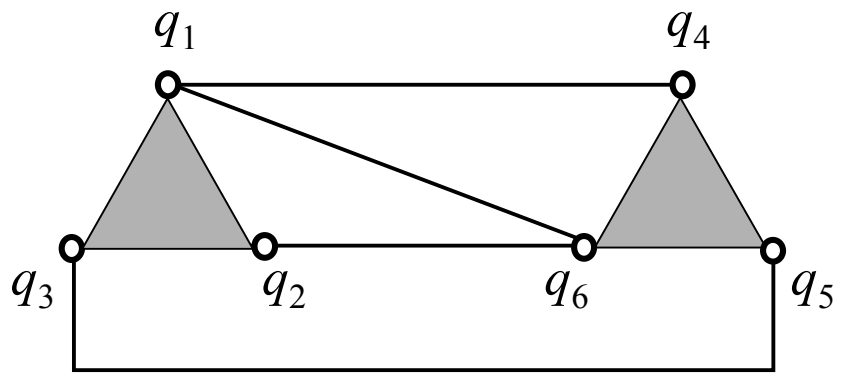
\end{document}